\documentclass[twocolumn,superscriptaddress,aps,pra,nobibnotes,numerical]{revtex4-1}

\usepackage{lipsum,capt-of,graphicx}
\usepackage[squaren]{SIunits}
\usepackage{verbatim}
\usepackage{mathrsfs}
\usepackage{comment}
\usepackage{braket}
\usepackage{bbold}
\usepackage{subfigure}
\usepackage{dcolumn}
\usepackage{amsmath}
\usepackage{hyperref}
\usepackage{enumerate}
\usepackage[11pt]{moresize}
\usepackage[normalem]{ulem}
\usepackage{xcolor}
\usepackage{color}
\usepackage{times}
\usepackage{amsmath}
\usepackage{amssymb}
\usepackage{booktabs}

\renewcommand{\thesubfigure}{\thefigure.\arabic{subfigure}} \makeatletter
\renewcommand{\p@subfigure}{}
\renewcommand{\@thesubfigure}{\thesubfigure.\,\,\hskip\subfiglabelskip} \makeatother

\begin{document}

\title{Nonequilibrium gas--liquid transition in the driven-dissipative
photonic lattice}

\author{Matteo Biondi} 
\affiliation{Institute for Theoretical Physics, ETH Zurich, 8093 Z\"urich, Switzerland}
\author{Gianni Blatter}
\affiliation{Institute for Theoretical Physics, ETH Zurich, 8093 Z\"urich, Switzerland}
\author{Hakan E. T\"ureci}
\affiliation{Department of Electrical Engineering, Princeton University, 08544 Princeton, 
New Jersey, USA}
\author{Sebastian Schmidt}
\affiliation{Institute for Theoretical Physics, ETH Zurich, 8093 Z\"urich, Switzerland}

\begin{abstract}
We study the nonequilibrium steady state of the driven-dissipative Bose-Hubbard
model with Kerr nonlinearity. Employing a mean-field decoupling for the
intercavity hopping $J$, we find that the steep crossover between low and
high photon-density states inherited from the single cavity transforms into a
gas--liquid bistability at large cavity-coupling $J$. We formulate a van der
Waals like gas--liquid phenomenology for this nonequilibrium setting and
determine the relevant phase diagrams, including a new type of diagram where
a lobe-shaped boundary separates smooth crossovers from sharp, hysteretic
transitions.  Calculating quantum trajectories for a one-dimensional system,
we provide insights into the microscopic origin of the bistability.
\end{abstract}

\maketitle

\section{Introduction}

The Bose-Hubbard Hamiltonian, describing strongly interacting bosons hopping
on a lattice, defines one of the fundamental model systems of condensed matter
physics and quantum optics. Its equilibrium phase diagram is characterized by
a lobe structure that results from a commensuration effect at integer particle
filling per site \cite{Fisher1989}. The phase boundary separating superfluid
from Mott-insulating phases is well understood \cite{Fisher1989,Jaksch1998}
and has been observed in landmark experiments on cold gases
\cite{Greiner2002,Bakr2010}. Coming to grips with Bose-Hubbard physics remains
a challenge in the photonic arena, where drive and dissipation are central to
the nonequilibrium model describing a lattice of nonlinear coupled cavities
\cite{Hartmann2006}. In this paper, we employ a mean-field decoupling in the
inter-cavity hopping $J$ on top of the exact single-cavity solution
\cite{Drummond1980}. We establish a van der Waals like gas--liquid
phenomenology and propose a new type of nonequilibrium phase diagram that
addresses the {\it nature} of the transition between phases. We find a
boundary that separates smooth from hysteretic transitions between photonic
gas and liquid phases and exhibits a pronounced quantum commensuration effect
in the cavity photon number.  Quantum trajectories for a chain of cavities
show that local density-fluctuations in individual cavities at small $J$
transform into collective super-cavity fluctuations and intermittent light
bursts when cavities become strongly coupled at large $J$.

The challenge in understanding the driven lattice roots in the complexity of the
single nonlinear cavity with its distinct low and high photon-density states
separated by a steep crossover. The experimental observation of bistability
between such states in a nonlinear optical device \cite{Gibbs1976} triggered a
vast amount of theoretical work \cite{Carmichael1977,Bonifacio1978,
Drummond1980,Gang1990,Bishop2010,Weimer2015,Carmichael2015,MendozaArenas2016,Casteels2016,Minganti2016}. 
Similar hysteretic cycles have been measured in different platforms and utilized in
the context of switching and amplification, e.g., with Josephson junctions
\cite{Siddiqi2004} and exciton-polaritons in semiconductor microcavities
\cite{Bajoni2008,Amo2010,Paraiso2010,Rodriguez2016_2}. While such
single-cavity physics is now well understood, new research perspectives are
being developed to explore bistable behavior in extended systems
\cite{Eichler2014_2,Rodriguez2016}, where the photon hopping $J$ between
different cavities competes with the on-site nonlinearity $U$.

Early work on photonic lattices described an (artificial) equilibrium setting
with a chemical potential for polaritons
\cite{Greentree2006,Angelakis2007,Rossini2007,Aichhorn2008,Schmidt2009,
Schmidt2010}, exhibiting close similarities in its phase diagram with that of
the massive Bose-Hubbard model \cite{Fisher1989}. Furthermore, a proper
initialization of the photonic lattice \cite{Hartmann2006}, e.g., with an
appropriate pump-pulse \cite{Tomadin2010_2}, provided signatures for a
superfluid--insulator phase transition in a dissipative cavity lattice.
Quite different physics emerges, however, when the cavities are coherently
driven, breaking the $U(1)$ symmetry explicitly. In this case, a mean-field
theory predicts a bistability that takes the array's state abruptly from low-
to high-density phases and vice versa, as was noted for the
Jaynes-Cummings-Hubbard model \cite{Nissen2012} and similarly for the
Bose-Hubbard model with Kerr nonlinearity \cite{Boite2013,Boite2014}. On the
experimental front, a bistable behavior has recently been observed on a large
one-dimensional circuit QED array \cite{Fitzpatrick2017}, further motivating a
deeper understanding of bistable behavior in large lattices.

\begin{figure}[b]
\includegraphics[width=0.475\textwidth]{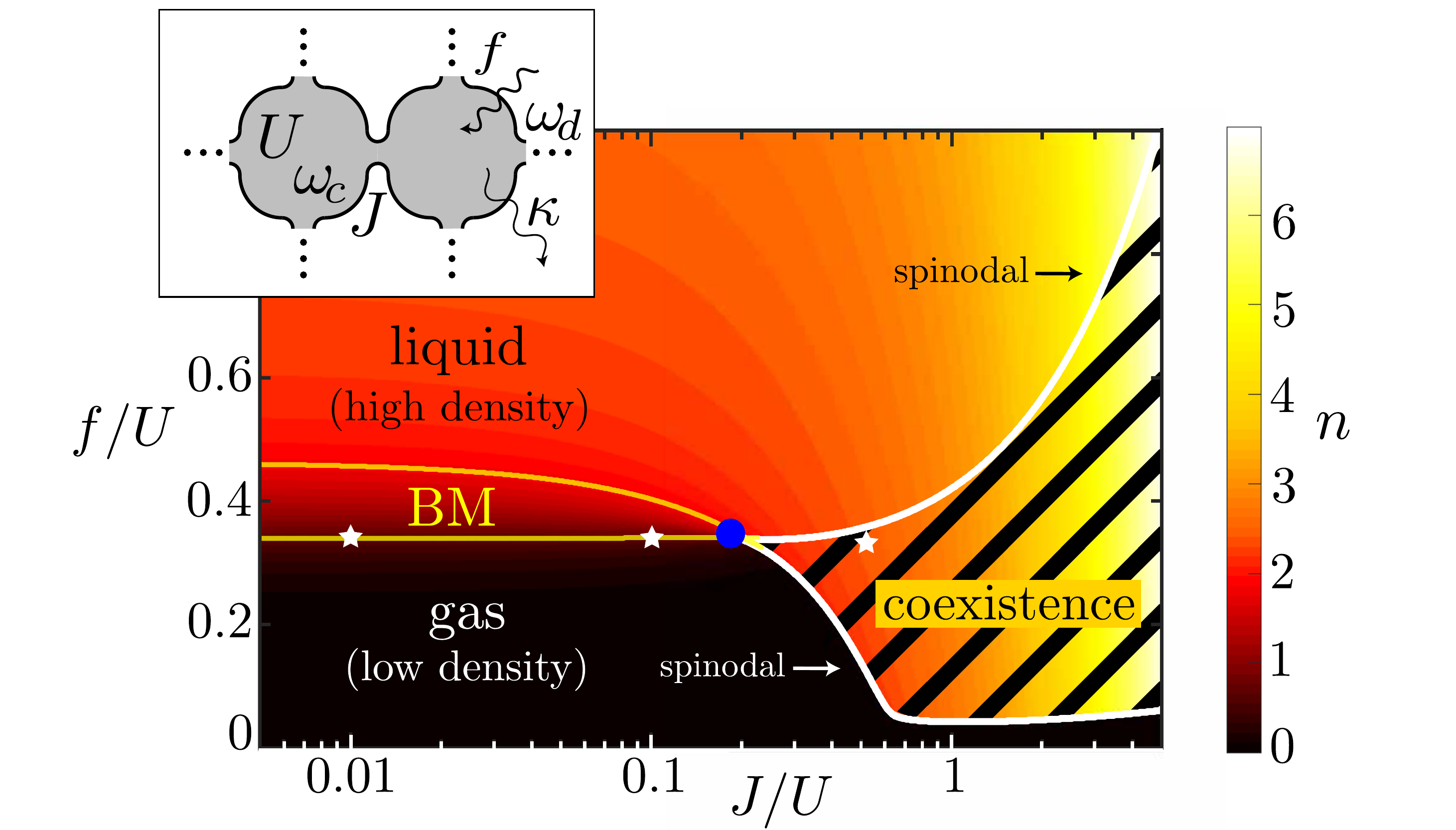}
\caption{(color online). Mean-field phase diagram of an array of nonlinear
cavities with interaction $U$ and loss $\kappa$, pumped with amplitude $f$ at
a frequency $\omega_d$ detuned from the cavity frequency $\omega_c$ by $\Delta
= \omega_d - \omega_c$, see top-left inset. Photons tunnel to neighboring
cavities with amplitude $J$. The photon density $n$ at the 4-photon resonance
$1 + 2\Delta/U = 4$ is shown as a function of the dimensionless parameters
$f/U$ and $J/U$ for small dissipation $\kappa=U/20$.  The smooth gas--liquid
crossover at small $J/U$ exhibits bimodality (BM region, yellow lines) in the
photon number distribution, and gives way to a hysteretic transition at $J_c
\approx 0.18\,U$ (dot), opening a coexistence region of gas and liquid at $J >
J_c$ (stripes; colors refer to densities in gas and liquid). The resulting
underdriven liquid and overdriven gas phases terminate at the spinodal lines
(white), which smoothly extend the lines bounding the bimodal region at small
$J$. The stars mark the location of the quantum trajectory results in
Fig.~\ref{fig:quant_traj}. \label{fig:g-l}}
\end{figure}

Despite such promising results, no clear view has emerged so far regarding the
nature and shape of the nonequilibrium diagram and its relation to the
equilibrium Bose Hubbard model, if there exists any at all. In particular, the
variety of tunable parameters and drive schemes makes the study of the
nonequilibrium photonic lattices a challenging problem. While the hopping $J$
is the obvious choice to track intercavity correlations, the replacement of
the chemical potential $\mu$ of the Bose-Hubbard model is less clear. It turns
out, that driving the cavities at a frequency $\omega_d$ different from the
cavity frequency $\omega_c$, the detuning $\Delta=\omega_d-\omega_c$ allows to
take the system in and out of many-photon resonances that assume a similar
role as the integer site-occupation in the Mott lobes, motivating its use in
replacing $\mu$.  Finally, imposing a coherent drive $f$, it is the
gas--liquid transition with its van der Waals type phenomenology rather than
the insulator--superfluid transition that plays the central role in this
system.
\begin{figure}[b]
\includegraphics[width=0.35\textwidth]{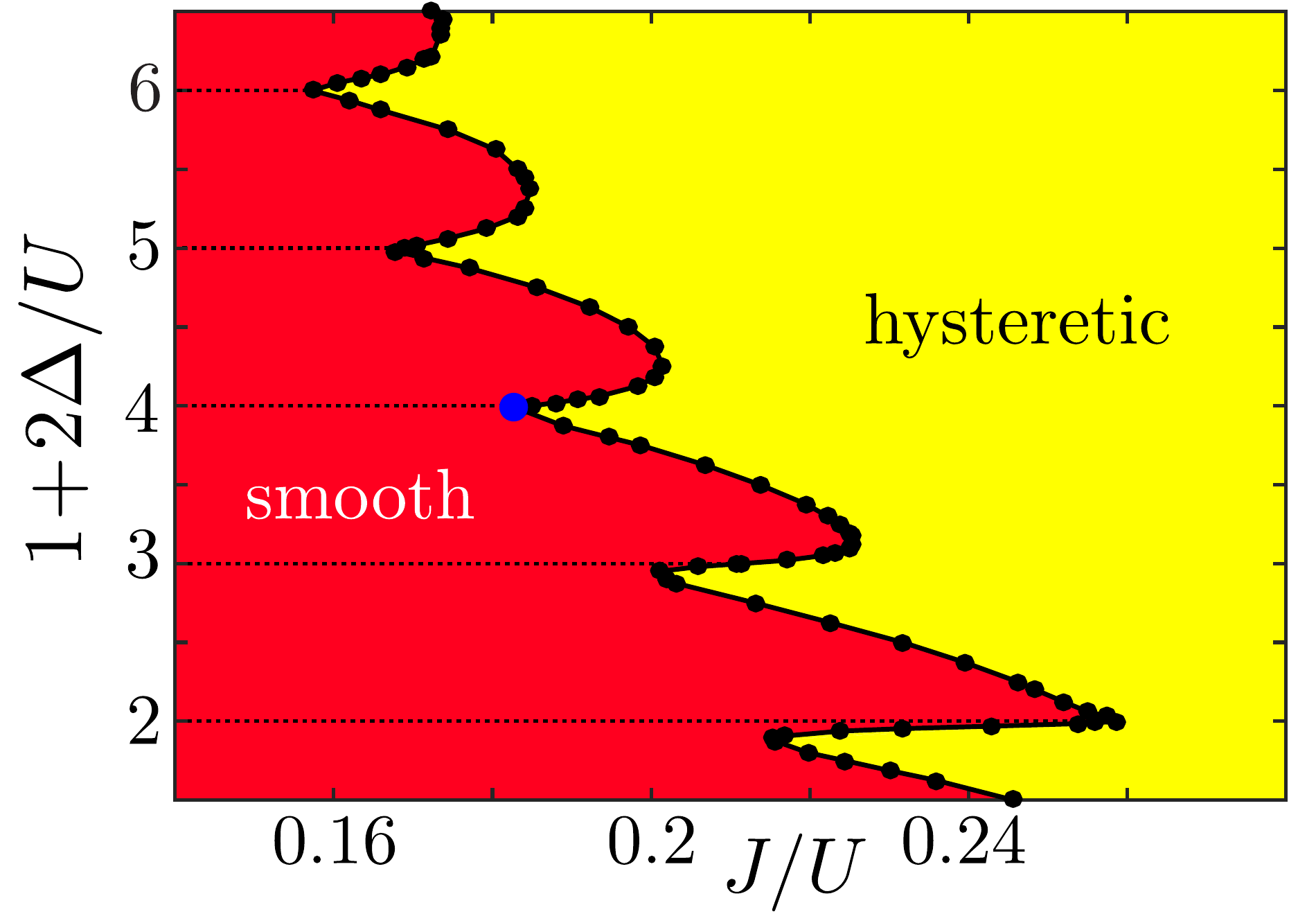}
\caption{(color online). Mean-field phase diagram displaying the nature of the gas--liquid
transition in the driven-dissipative photonic lattice.  Plotting the
dimensionless detuning $2\Delta/U$ versus hopping $J/U$ at small dissipation
$\kappa=U/20$, we show the boundary separating smooth from hysteretic
gas--liquid transitions as driven by increasing the pump amplitude $f$.
Distinct lobes appear between successive $m$-photon resonances of the
individual cavities, i.e., when $1+2\Delta/U=m$ assumes an integer value, thus
featuring a similar commensuration effect as the equilibrium Bose-Hubbard
model. Going to small $\Delta/U$ or very small dissipation $\kappa$,
instabilities show up in the mean-field analysis, see also
Refs.~\cite{Boite2013,Wilson2016}. Numerical errors are of order the size of
the points.\label{fig:m-J}}
\end{figure}

In our analysis, we make use of a mean-field decoupling scheme in the hopping
$J$. Such a mean-field description has been very successful in predicting the
qualitative features of the equilibrium phase diagram of the Bose-Hubbard model,
motivating its use for the investigation of our nonequilibrium setting as
well. The results of our analysis are expressed in two phase diagrams.
Fig.~\ref{fig:g-l} shows how the gas--liquid transition as driven by the
coherent pump amplitude $f$ changes from a {\it steep crossover} inherited
from the single cavity at small $J$ to a first-order type {\it hysteretic or
bistable} transition at large $J$. The termination of the hysteretic behavior
upon decreasing $J$ then defines a critical end-point to a first-order like
transition in the $f$--$J$ diagram at fixed detuning $\Delta$.  In
Fig.~\ref{fig:m-J}, we track the location of this critical end-point in a
$\Delta$--$J$ diagram and find a boundary with characteristic lobes appearing
between successive $m$-photon resonances of the individual cavities where
$1+2\Delta/U=m$ assumes integer values. This boundary separates regions where
the gas--liquid transition is smooth (small $J/U$) from regions where
bistability governs the lattice's behavior as the pump amplitude $f$ is tuned
across the transition. Contrary to conventional phase diagrams describing
transitions between phases, our $\Delta$--$J$ phase diagram addresses the {\it
nature} of the transition, smooth versus hysteretic, as the system parameters
are changed.

\section{Driven-Dissipative Bose-Hubbard Model}

We consider the driven-dissipative Bose-Hubbard (BH) model, describing photons
hopping on a lattice of nonlinear cavities, pumped and lossy. The Hamiltonian
($\hbar=1$) reads
\begin{equation}
\label{h_BHM}
   H = \sum_i h^{\scriptscriptstyle \rm BH}_i + \frac{1}{z}
   \sum_{ \langle ij \rangle } J_{ij} a^\dagger_i a_j
\end{equation}
with $h^{\scriptscriptstyle \rm BH}_i = - \Delta\,n_i + Un_i(n_i - 1)/2 +
f(a_i + a_i^\dagger)$, the bosonic operators $a_i$ and the number operators
$n_i=a^\dagger_ia_i$.  Each site $i$ is coherently pumped with strength $f$ as
described by the last term in $h^{\scriptscriptstyle \rm BH}_i$. In a frame
rotating with the drive frequency $\omega_d$, the cavity frequency is
renormalized to $\Delta=\omega_d-\omega_c$, while $U$ is the local Kerr
nonlinearity. The second term in $H$ describes the hopping to $z$
nearest-neighbor cavities with amplitude $J_{ij}=-J$; the factor $1/z$ in
Eq.~\eqref{h_BHM} ensures a bandwidth $2J$ independent of $z$ and guarantees a
regular limit $z\to\infty$ where mean-field theory becomes exact.  The
dissipative dynamics for the density matrix $\rho$ is determined by the
Lindblad master equation
\begin{equation} \label{lindblad_full}
   \dot{\rho} = -i[H,\rho] + \frac{\kappa}{2} 
   \sum_i (2a_i\rho a_i^\dagger-a_i^\dagger a_i\rho-\rho a_i^\dagger a_i),
\end{equation}
with the photon decay rate $\kappa$.  Models of this type can be realized in
quantum engineered settings using superconductor-
\cite{Houck2012,Schmidt2013_2,Leib2014} and semiconductor technologies
\cite{Carusotto2013,Baboux2015}.

\subsection{Single Cavity}

\begin{figure}[t]
\includegraphics[width=0.4925\textwidth]{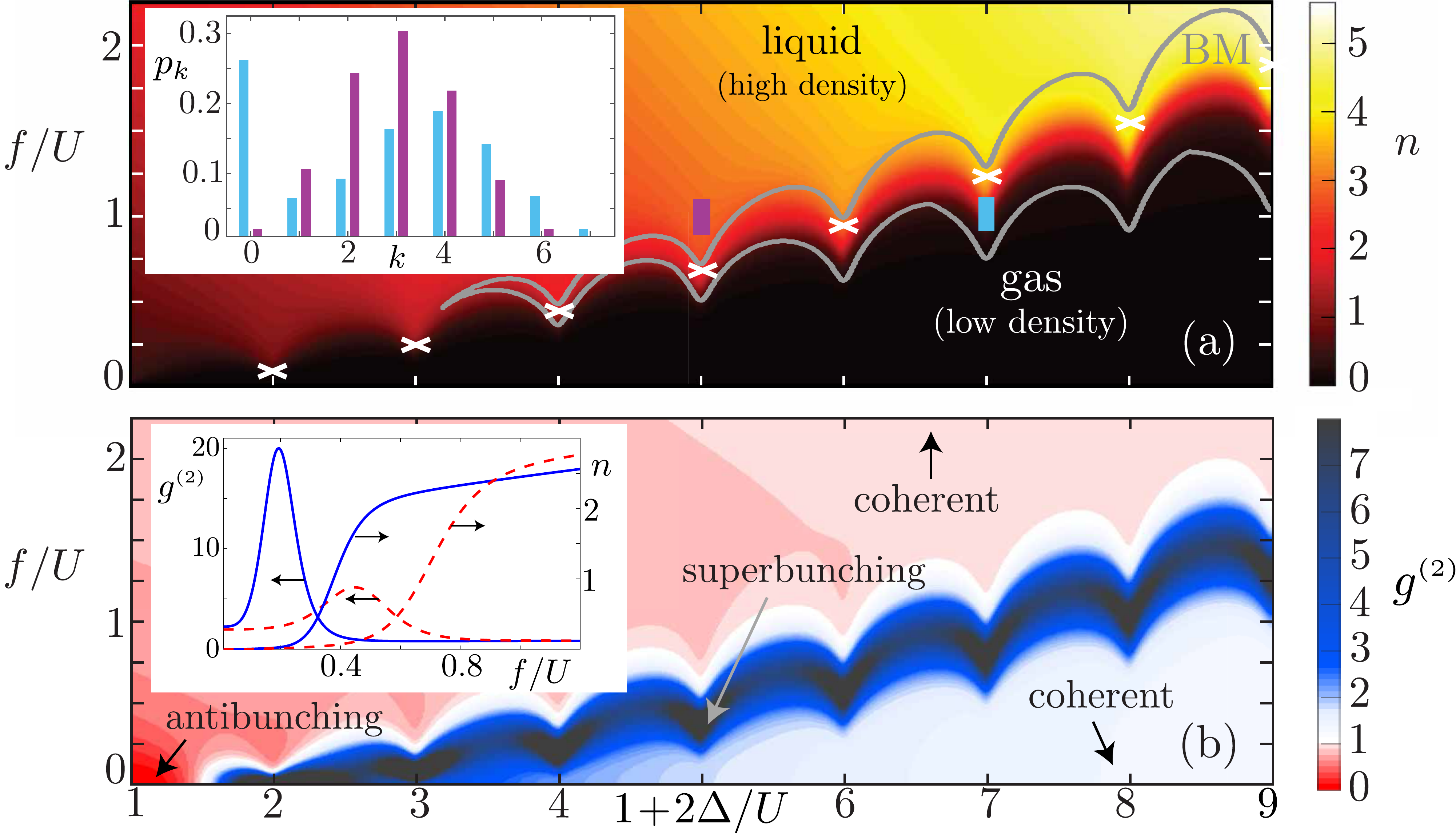}
\caption{(color online). Density $n$ (a) and second-order coherence
$g^{\scriptscriptstyle (2)}$ (b) as a function of drive detuning $2\Delta/U$
and drive strength $f/U$ for a single cavity as obtained from the exact
solution \cite{Drummond1980} of equation~\eqref{lindblad_full} with $J=0$ and
$\kappa = U/20$. The modulated grey lines labelled `BM' encompass the bimodal
regime.  The inset displays the photon number distribution $p_k$ at the two
bars marked in the main panel. The white crosses mark the onset
$f^\mathrm{sc}_{\!\scriptscriptstyle\times}$ of the liquid phase as defined by
the condition of unit compressibility $K=1$.  The correlator
$g^{\scriptscriptstyle(2)}$ illustrates the phases' coherent nature, while the
crossover is characterized by superbunching, see also
Ref.~\cite{Fitzpatrick2017}.  The bottom inset displays the density $n$ and
correlator $g^{\scriptscriptstyle (2)}$ evaluated at fixed detunings
$2\Delta/U=3,~3.5$ (solid, dashed).\label{sc_pdiag}} \end{figure}

The driven-dissipative single cavity (i.e., equation~\eqref{lindblad_full}
with $J = 0$) has been solved exactly by Drummond and Walls
\cite{Drummond1980} and the results are summarized in Fig.~\ref{sc_pdiag}.
The diagram in Fig.~\ref{sc_pdiag}(a) exhibits two states or phases
characterized by low and high photon-densities $n = \langle a^\dagger
a\rangle$. The crossover from the low- (gas) to the high-density (liquid)
phase is driven via increasing the pumping amplitude $f$ and exhibits
bimodality in the photon number distribution $p_k$, see also
Ref.~\cite{Fink2016}.  We estimate the location of the crossover line by
comparing terms in the Hamiltonian $h^{\scriptscriptstyle \rm BH}$, generating
scalings $n\sim(f/\Delta)^2$ at small drive $f$ (gas-phase) and
$n\sim(f/U)^{2/3}$ in the liquid phase at large $f$ where the interaction $U$
dominates. The crossover between the two regimes appears at $n\sim\Delta/U$
and defines the crossover line
$f^\mathrm{sc}_{\!\scriptscriptstyle\times}/U\sim(\Delta/U)^{3/2}$. We obtain a
more quantitative result from the exact solution \cite{Drummond1980} at weak
dissipation $\kappa/U\ll1$: with the compressibility
$K=1+n(g^{\scriptscriptstyle (2)}-1)$ dropping below unity upon entering the
liquid phase ($g^{\scriptscriptstyle (2)}=\langle a^\dagger a^\dagger
aa\rangle/n^2$ the second-order coherence), the condition $K = 1$ provides the
result $f^\mathrm{sc}_{\!\scriptscriptstyle\times}/U \approx (m/2e)^{3/2}
(m\kappa/U)^{1/m}$ at the $m$-photon resonance $2\Delta=(m-1)\,U$ (where the
energy of $m$ photons outside and inside the cavity match up), which agrees
(up to a numerical coefficient) with our previous estimate at large $m$.

The interaction leads to an intermediate plateau in the liquid phase with
density $n\approx\Delta/U$, see inset in Fig.~\ref{sc_pdiag}(b) (the $1/2$
reduction in $n$ with respect to $m$ is a saturation effect \cite{Biondi2015}). 
The transition to the liquid is helped when the drive frequency is resonant with the
$m$-photon state of the cavity at $2\Delta/U =(m-1)$, yielding the modulation
of the crossover line in Fig.~\ref{sc_pdiag}, see also
Ref.~\cite{Carmichael2015}.  The low- and high-density phases are well
described by coherent states (except for small $f$ and $\Delta$) as quantified
by the correlator $g^{\scriptscriptstyle (2)}$. The crossover in between is
characterized by large density fluctuations and superbunching, see
Fig.~\ref{sc_pdiag}(b).

\subsection{Cavity Lattice}

We now combine cavities into a lattice and increase the intercavity hopping
$J$. We solve for the non-equilibrium steady state $\dot\rho = 0$ of the photonic
lattice by reducing the task to a single-site problem via a mean-field
decoupling of the hopping term \cite{Tomadin2010_2,Tomadin2011} in equation
\eqref{lindblad_full}, i.e, $a^\dagger_i a_j \approx  \langle a^\dagger_i
\rangle a_j + a^\dagger_i \langle a_j \rangle - \langle a_i^\dagger \rangle
\langle a_j \rangle$; the same decoupling has been used in the equilibrium
model \cite{Fisher1989} and provided correct qualitative results for the phase
diagram. Alternatively, the same approximation can be obtained from an
expansion of the lattice density matrix in inverse powers of the coordination
number $z$ \cite{Navez2010}; truncating the expansion at order unity is
equivalent to the mean-field decoupling of the hopping term and is exact in
the limit $z\rightarrow \infty$, i.e., large dimensions. We then obtain a
self-consistent equation \cite{Drummond1980,Boite2013} for the mean amplitude
$\langle a_i \rangle = \langle a \rangle$,
\begin{equation}
\label{eq:mf}
   \langle a \rangle = -\frac{2|\varphi_J|}{\delta} 
   \frac{_0F_2(;1+\delta,\delta^\ast;8|\varphi_J|^2)}
   {_0F_2(;\delta,\delta^\ast;8|\varphi_J|^2)},
\end{equation}
with the renormalized drive $\varphi_J=(f-J\langle a\rangle)/U$ depending on
$\langle a\rangle$, the dimensionless detuning $\delta=-(2\Delta+i\kappa)/U$
and the hypergeometric function $_0F_2(;a,b;z)$; the solution for $\langle
a\rangle$ provides direct access to the photon density $n= \langle a^\dagger_i
a_i\rangle = \langle a^\dagger a\rangle$ and higher-order correlators
\cite{Drummond1980}.  Eq.~\eqref{eq:mf} exhibits multiple solutions at large
hopping $J$. The location $J_c$ where these multiple solutions
first show up is our main interest here, since it describes the transition
from a {\it smooth} gas--liquid crossover in the density $n$ as observed in
the single cavity, to a {\it hysteretic} first-order type transition
characteristic of a strongly-coupled lattice system.

The driven Bose-Hubbard model involves the parameters $f$, $U$, $J$, and
$\Delta$, and it is the suitable choice within this set which brings forward
the properties of this system. In a first step, we fix the dimensionless
detuning $\Delta/U$ to the four-photon resonance at $1+2\Delta/U=4$ and
increase the drive $f/U$. This produces the gas--liquid phase diagram in
Fig.~\ref{fig:g-l}, where the density $n$ assumes the role of the order
parameter.  At small hopping $J/U < 0.18$, gas and liquid phases are separated
by a steep crossover with a bimodal distribution $p_k$ of photon numbers
inherited from the single cavity. The location of this crossover is well
described by the compressibility criterion $K=1$, resulting in a line
following accurately the upper boundary of the bimodal region in
Fig.~\ref{fig:g-l}; an approximation in the small-$\kappa$ limit
\cite{Boite2014} yields a linear dependence on $J$,
\begin{equation}
  f_{\!\scriptscriptstyle\times} \approx f_{\!\scriptscriptstyle\times}^\mathrm{sc} 
 (1 - 2J/U),
\label{pbound_ca}
\end{equation}
with $f^\mathrm{sc}_{\!\scriptscriptstyle\times}$ the single-cavity expression
derived with the same condition $K=1$. The smooth crossover between gas and
liquid phases ends at a `critical' value $J_c \approx 0.18\,U$ (blue
dot), corresponding to $f_c \approx 0.29\,U$, giving way to a
hysteretic transition at larger hopping $J/U$ that shows the signatures
typical of a van der Waals like gas--liquid transition \cite{Domb1996}: using this
terminology, we find two-phase coexistence bounded by spinodal lines at large
coupling $J$ that smoothly develop out of the bimodal lines at small coupling.
Similar results are obtained at different values of the misfit parameter
$\Delta/U$, but with a plateau at a suitably adapted photon density, $n
\approx \Delta/U$.

Evaluating the location of the critical point $J_c$ for different detunings
$\Delta/U$, we can plot a boundary separating smooth from hysteretic behavior
and arrive at a complete characterization of the system. We find a boundary
with a lobe-like structure that is commensurate with the $m$-photon resonances
at integer values of $1+2\Delta/U$, see Fig.~\ref{fig:m-J}, a result that has
been searched for in the past, but has remained elusive so far.

\section{Quantum Trajectories}

\begin{figure}[t] 
\includegraphics[width=0.48\textwidth]{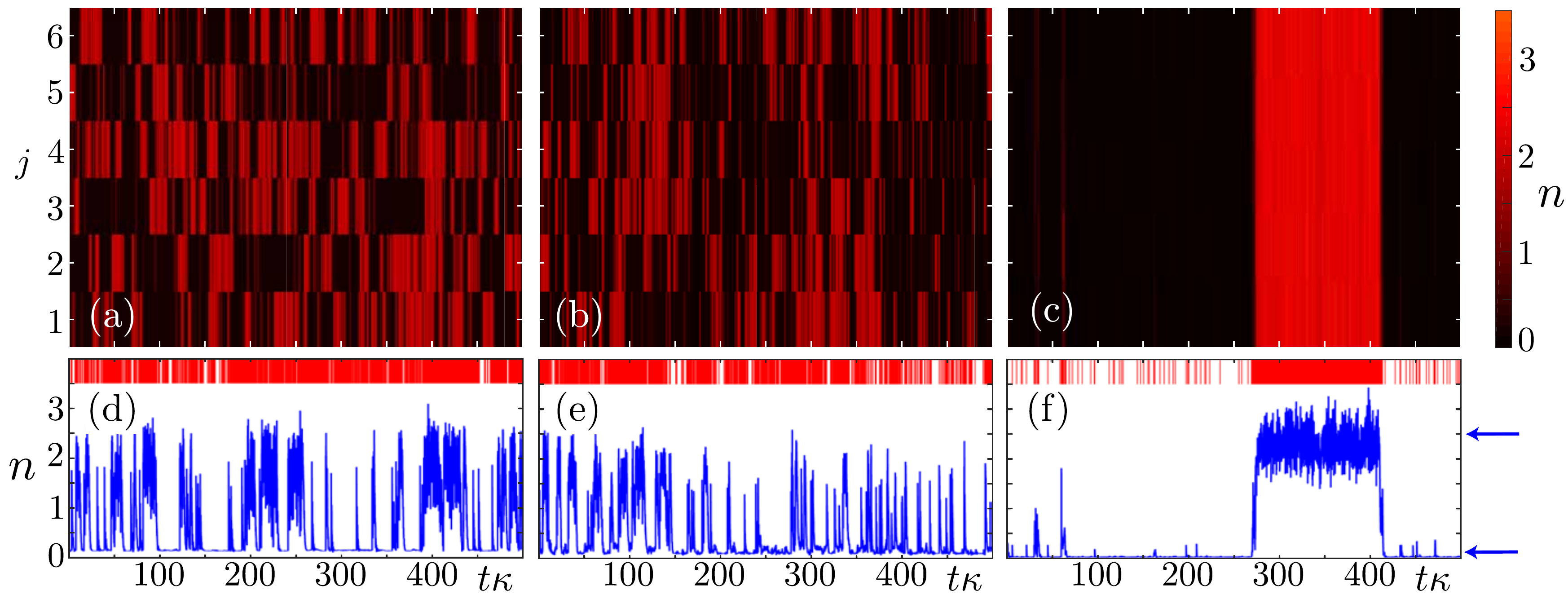}
\caption{(color online). Selected quantum trajectories for a 1D cavity array
with 6 sites.  The panels (a)--(c) show the photon density in color scale as a
function of time ($t\kappa$) and position (lattice site $j$) at fixed drive
strength $f/U = 0.35$ and for increasing hopping $J/U$ as marked with the
stars in the phase diagram of Fig.~\ref{fig:g-l}. At small hopping $J\ll J_c$,
the trajectories of different sites are uncorrelated (a), while for $J> J_c$,
the entire cavity array switches collectively between gas and liquid states
within the coexistence region of the mean-field diagram, see (c).  The panels
(d)--(f) show trajectories for a single lattice site $j=4$ as a function of
time, as taken from the respective top panels (a)--(c).  The vertical red bars
indicate the photon emissions from the lattice. For $J\ll J_c$ (d) each
individual cavity displays intermittency \cite{Plenio1998} (see also text) at
random times, yielding a constant photon emission from the array. For $J> J_c$
(f) the array behaves as a coherent super-cavity and a collective
intermittency is restored. The horizontal arrows mark the gas and liquid
mean-field values, showing that collective switching in panel (c) indeed
occurs between the mean-field densities. Panels (a), (b), (c): $J/U=0.01, 0.1,
0.5$. Other parameters are chosen as in Fig.~\ref{fig:g-l}.  Convergence of
the quantum trajectory results in the photon truncation parameter (cutoff) is
illustrated in Appendix \ref{appendix:convergence}.  \label{fig:quant_traj}}
\end{figure} 

In order to substantiate our results, we complete this study with a
microscopic view on the gas--liquid diagram in Fig.\ \ref{fig:g-l}. In
Fig.~\ref{fig:quant_traj}, we present simulation results of selected quantum
trajectories \cite{Dalibard1992,Carmichael1992} (see also the reviews
\cite{Plenio1998,Daley2014} and the Appendix \ref{appendix:trajectories} for
further information) for a chain of 6 nonlinear cavities in one dimension (1D)
with periodic boundary conditions.  At small values of $J$, the cavities
switch individually between gas and liquid states, see panel (a), with a
rapidly growing weight of the liquid when $f$ is tuned across
$f_{\!\scriptscriptstyle\times}/U \sim (\Delta/U)^{3/2}$. As $J/U$ is
increased within the bimodal region, the fluctuations become correlated and
extended super-cavities are formed, see panel (b). Increasing $J$ further
across $J_c$, the entire strongly-coupled array switches collectively as
illustrated by the appearance of pronounced stripes in panel (c) of Fig.\
\ref{fig:quant_traj}, with switching times largely exceeding those of the
individual cavities.  In an infinite system, we then expect a second-order
transition with a diverging correlation length to appear as $J$ is increased
towards $J_c$ in the bimodal strip. This hypothesis is supported by
simulations exhibiting a rapid increase of the collective switching time with
system size, suggesting a closing of the Liouvillian gap in the thermodynamic
limit (see Appendix \ref{appendix:switching}), and invites for further
exploration, also with a view on the role of lattice dimensionality
\cite{FossFeig2017}. On the other hand, increasing the drive $f$ at fixed
coupling $J > J_c$, we expect a first-order type behavior with nucleation of
extended liquid phases in the gas and vice versa on decreasing $f$. The
intermittent light bursts appearing in the hysteretic regime, cf.\ the red
photon emission processes shown in Fig.\ \ref{fig:quant_traj}(f), naturally
show up in the context of dynamical phase transitions \cite{Ates2012} and can
serve as an experimental probe of the hysteretic behavior
\cite{Fitzpatrick2017}.  We note that quantum trajectories obtained in related
models, assemblies of Rydberg atoms \cite{Ates2012,Lee2012} and spin-1/2 $XY$
models \cite{Wilson2016}, also exhibit collective switchings between phases
but do not show individual fluctuations with a transition between the two
behaviors.

In comparing the physics of the two versions of the Bose-Hubbard model,
equilibrium versus coherently-driven--dissipative, we note that the former is
characterized by a phase boundary $J_c (\mu)$ describing a spontaneous
breaking of $U(1)$ symmetry, while the latter exhibits the phenomenology of a
tunable van der Waals type gas--liquid transition. In particular, in the
coherently driven system, the $U(1)$ symmetry is explicitly broken and the
interesting feature is the transformation of a smooth crossover into a
hysteretic transition involving local (at small $J$) or collective (at large
$J$) temporal fluctuations of low- and high-density phases. In spite of the
differences between the two phenomenologies, both phase boundaries $J_c(\mu)$
and $J_c(\Delta)$ exhibit a particle commensuration effect resulting in a
lobe-like structure. In the equilibrium situation, the superfluid phase is
favored whenever the chemical potential $\mu$ allows for two different
particle numbers, while in the driven Bose-Hubbard model, a detuning $\Delta$
matching a many-photon resonance in each cavity facilitates their
synchronization and thereby triggers collective jumps between gas- and liquid
photonic phases. This can be understood as a variation of Le Chatelier's
principle stating that the system reacts to a disturbance, here a change in
$\mu$ or $\Delta$, by favoring the corresponding phase, superfluid when
particle number becomes undefined and intermittent light bursts when
approaching a resonance.

\section{Summary and Conclusions}

Summarizing, we have presented a mean-field analysis of the driven-dissipative
Bose-Hubbard model describing a lattice of coupled nonlinear cavities.
Inspired by the exact single-cavity solution with its crossover between low-
and high-density phases, we have established a van der Waals type gas--liquid
phenomenology for the driven photonic Bose-Hubbard model featuring a change
from smooth to hysteretic transition upon increasing the coupling $J$ beyond
critical. A quantum-trajectory analysis shows that the bistable region
involves collective switching between gas- and liquid phases triggering bursts
of light. Choosing the correct representation in parameter space, both
equilibrium and driven phase diagrams exhibit boundaries with a lobe-like
structure that originates from a resonance condition in the on-site
Hamiltonian. We expect that models with a similar on-site nonlinearity, e.g.,
the Jaynes-Cummings-Hubbard model \cite{Greentree2006,Nissen2012} will exhibit
an analogous phase diagram, while models of similar kind, e.g., assemblies of
Rydberg atoms and spin-1/2 systems
\cite{MendozaArenas2016,Ates2012,Lee2012,Wilson2016}, will benefit from the
insights obtained in this paper.  Our results clarify a long-standing problem
on the nature and shape of the phase diagram of the driven Bose-Hubbard model
and guide new experiments on photonic arrays.

\begin{acknowledgments}

We thank H.-P.\ B\"uchler, T.\ Esslinger, S.\ Huber, O.\ Zilberberg, and W.\
Zwerger for discussions and acknowledge support from the Swiss National
Science Foundation through an Ambizione Fellowship (SS) under Grant No.\
PZ00P2\_142539, the National Centre of Competence in Research `QSIT--Quantum
Science and Technology' (MB) and the US Department of Energy, Office of Basic
Energy Sciences, Division of Material Sciences and Engineering under Award
No.\ DE-SC0016011 (HET).

\end{acknowledgments}

\appendix

\section{Quantum trajectory approach \label{appendix:trajectories}}

In here, we briefly summarize the quantum trajectory algorithm introduced in
the Refs.~\cite{Dalibard1992} and \cite{Carmichael1992} and well documented in 
reviews, see, e.g., Refs.~\cite{Plenio1998} and
\cite{Daley2014}. The algorithm is used to describe open quantum
systems whose dynamics is described by a master equation in Lindblad form, as
Eq.~\eqref{lindblad_full} in the main text. The quantum trajectory method is, (i) numerically
advantageous with respect to the direct integration of the master equation,
and (ii) can provide further insight into the dynamical behavior of the system
due to the stochastic nature of the trajectories. The algorithm
stochastically propagates the wavefunction $\ket{\psi(t)}$ under the
non-hermitian Hamiltonian
\begin{equation}
H_{\rm eff} = H - i\frac{\kappa}{2}
\label{eq:h_eff}
\sum_j a^\dagger_j a_j
\end{equation}
with the photon decay rate $\kappa$. The Hamiltonian $H$ of Eq.~\eqref{eq:h_eff}, the
density operator $n_j = a^\dagger_j a_j$ and the photon operator $a_j$ have
been introduced in Eq.~\eqref{h_BHM} of the main text. The algorithm can be summarized
as follows. If in the time interval $[t,t+dt]$ the cavity at site $j$ emits a
photon, the wavefunction collapses to $\ket{\psi(t+dt)} = a_j \ket{\psi(t)}$,
while, if no photon is emitted, $\ket{\psi(t+dt)} = (1  - i H_{\rm eff}\,dt)
\ket{\psi(t)}$. Which of these events occurs depends on the photon density
$n_j(t) = \braket{\psi(t) | n_j | \psi(t)}$ and is determined stochastically
by comparison with a random number. This process can be understood as the
measurement of the system by the environment. This follows from the fact that
information is gained also when no photon is emitted. After normalizing the
wavefunction, the stochastic evolution continues with the next time step till
the \emph{trajectory} is complete. In practice, variants of the algorithm of
higher order in the time step $dt$ are used \cite{Daley2014}.

The quantum trajectory algorithm is numerically advantageous with respect to
the direct integration of the master equation, since it is based on
propagating the wavefunction instead of the density matrix; furthermore,
different trajectories are independent and can thus be propagated in parallel.
The average over different stochastic evolutions is equivalent to the density
matrix dynamics as determined by the Lindblad master equation given by Eq.~(2)
of the main text. Furthermore, in the single trajectories fundamental
information on the behavior of the system is revealed.

\section{Convergence in the photon truncation parameter \label{appendix:convergence}}

\begin{figure}[t]
\centering
\includegraphics[width=0.375\textwidth]{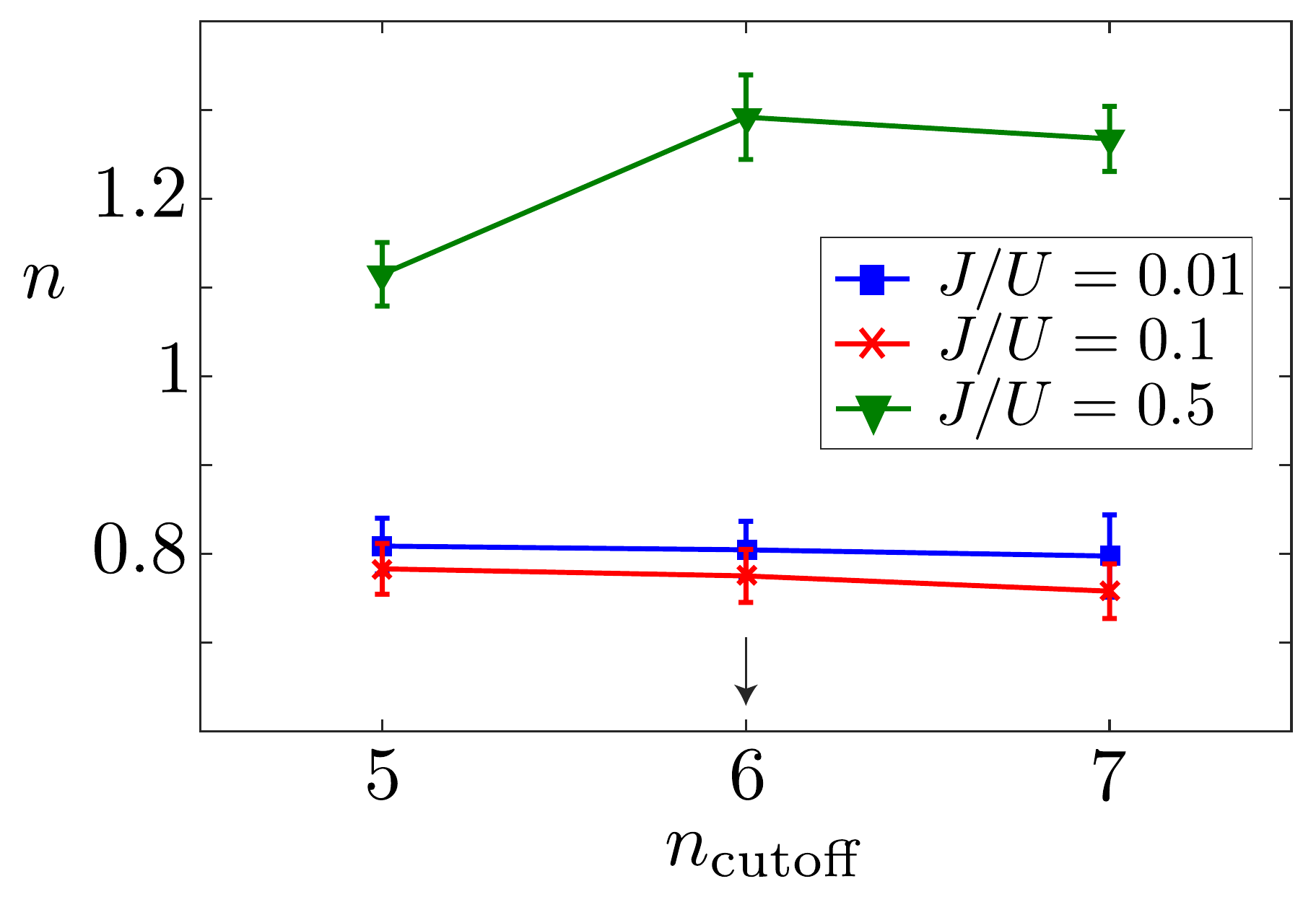}  
\caption{(color online). Convergence plot of the average photon density $n$ (see Eq.~\eqref{eq:av_dens}) in the steady state for various values of hopping strengths $J/U$ as calculated with quantum trajectories. The average density $n$ is shown as a function of the photon number truncation parameter $n_{\rm cutoff}$ for a lattice of $N=5$ sites with PBC. The vertical bars denote one standard deviation (see text). Other parameters as in Fig.~\ref{fig:quant_traj} of the main text. The vertical arrow indicates the cutoff value $n_{\rm cutoff} = 6$ used in Fig.~\ref{fig:quant_traj} of the main text, for an array of $N=6$ sites with PBC. \label{fig1_App}}
\end{figure}

To obtain Fig.~\ref{fig:quant_traj} in the main text, we employ a 5-th order Runge-Kutta (built
in the Matlab routine \emph{ode45}) to simulate the stochastic evolution as
outlined in Section 3.5 of Ref.~[\onlinecite{Daley2014}]. For the quantum
trajectories displayed in Fig.~\ref{fig:quant_traj} of the main text, up to 6 photons per cavity
are admitted, resulting in a Hilbert space of $7^6 = 117649 \approx 2^{17}$ states. Fig.~\ref{fig1_App} shows the convergence of the average photon density as a function of 
the photon number truncation parameter $n_{\rm cutoff}$ for different $J/U$ values for a lattice of $N=5$ sites with periodic boundary conditions (PBC). 
The average photon density is defined as
\begin{equation}\begin{split}
n  & = \langle\langle\langle n \rangle_{\rm time} \rangle_{\rm sites}\rangle_{\rm traj} \\
& = \frac{1}{N_{\rm traj}N_{\rm sites} N_{\rm t\text{-}steps}} \sum_{r=1}^{N_{\rm traj}}  \sum_{j=1}^{N_{\rm sites}}   \sum_{t=t_0}^{N_{\rm t\text{-}steps}} n_{j,r,t}.
\label{eq:av_dens}
\end{split}
\end{equation}
In the definition above, the average is taken first over time for $t \ge t_0$, with $t_0 \gg 1/\kappa$ such that a steady state is reached; the resulting density is averaged over different sites in the lattice and finally an average over the results obtained through independent trajectories is performed. The squared deviation from the mean (variance) is propagated according to the standard prescriptions of error propagation, yielding a final standard deviation $\sigma_n$. In the coexistence region of the mean-field (see main text) where the different sites in the array are correlated, only a specific site $j=4$ is considered and the average over different sites is discarded. In our convergence simulations, $N_{\rm t\text{-}steps} \approx 10^4$, $N_{\rm traj} = 100$ and $N_{\rm sites} = N = 5$. At small $J/U$ (blue and red symbols) we note that already $n_{\rm cutoff} = 5$ provides a good approximation. At larger hopping strengths (green symbols) we find that a larger cutoff is needed to reach convergence within one standard deviation.

\begin{figure*}[t]
\centering
\includegraphics[width=0.75\textwidth]{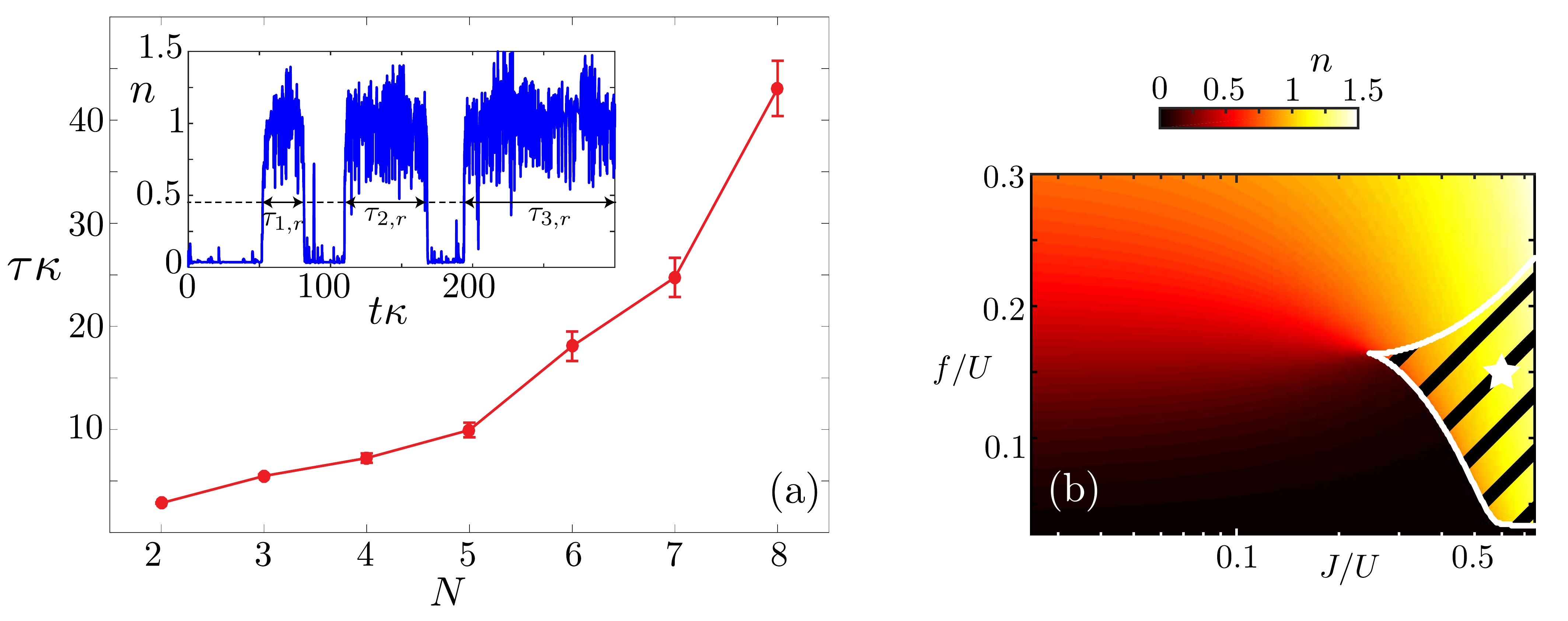}  
\caption{(color online). Collective time spent in the liquid phase $\tau$ (see Eq.~\eqref{eq:av_tau} and main text) in the steady state as calculated with quantum trajectories (QT) as a function
of the size $N$ of a one-dimensional array with PBC (a). The inset displays a sample trajectory for $N=8$ exhibiting 3 separate periods where the system dwells in the liquid phase (see also text). In order to extract $\tau_{s,r}$ (see text), an arbitrary threshold $n = 0.4$ is used to separate gas and liquid phases and density fluctuations on a scale smaller than $dt = 2/\kappa$ are neglected. We have checked that the trend in the results (exponential-like increase of $\tau$ with $N$) is invariant with respect to these choices. The trajectory results are shown for a set of parameters within the coexistence region of the mean-field (MF), $f/U = 0.15$, $J/U=0.6$, as indicated by the white star in (b). The detuning value $\Delta = \omega_d - \omega_c$ (detuning between the drive frequency and the cavity frequency) is set to $1+2\Delta/U = 1.55$. For this choice of detuning, convergence of the quantum trajectory results in the photon truncation parameter $n_{\rm cutoff}$ is achieved for $n_{\rm cutoff} = 3$, much lower than $n_{\rm cutoff} = 6$ required for the detuning value $1+2\Delta/U = 4$ used in Figs.~\ref{fig:g-l} and \ref{fig:quant_traj} in the main text. The choice of a lower detuning $\Delta/U$ with respect to Fig.~\ref{fig:quant_traj} in the main text thus allows us to study larger system sizes. The dissipation strength is chosen as $\kappa/U = 20$, as in Figs.~\ref{fig:g-l}--\ref{fig:quant_traj} in the main text. \label{fig2_App}}
\end{figure*}

\section{Scaling of the collective switching time with system size \label{appendix:switching}}
In this section we focus on the coexistence region of the mean-field (see main text) and discuss the scaling of the collective switching time (see main text) with system size as calculated with quantum trajectories. To this end we consider the average time $\tau$ spent in the liquid phase; in order to extract $\tau$ from an ensemble of trajectories we first obtain
the average time spent in the liquid phase in a single trajectory and successively average the result over different trajectories, i.e.,
\begin{equation}
\tau =  \langle\langle \tau \rangle_{\rm time} \rangle_{\rm traj} = \frac{1}{N_{\rm traj}} \sum_{r=1}^{N_{\rm traj}}\frac{1}{N_{\rm liquid}(r)}  \sum_{s=1}^{N_{\rm liquid}(r)} \tau_{s,r}.
\label{eq:av_tau}
\end{equation}
In the definition above $\tau_{s,r}$ is the time spent in the liquid phase in trajectory $r$ and in period $s$; the number of separate periods where the system dwells in the liquid phase 
in each trajectory is denoted by $N_{\rm liquid}(r)$ and is trajectory-dependent. The squared deviation from the mean (variance) is propagated according to the standard prescriptions of error propagation, yielding a final standard deviation $\sigma_\tau$. In these simulations $N_{\rm traj} = 150$. Fig.~\ref{fig2_App}(a) shows $\tau$ \eqref{eq:av_tau} as a function of system size for a set of parameters within the coexistence region of the mean-field, see Fig.~\ref{fig2_App}(b). The inset in Fig.~\ref{fig2_App}(a) shows a sample trajectory characterized by 3 separate periods; when the system is still in the liquid phase at the end of the trajectory, the latter is considered as the end of the period. We find that the the time spent in the liquid phase increases rapidly with system size; this result is consistent with our hypothesis on the emergence of a transition in the thermodynamic limit characterized by a closing of the Liouvillian gap (see main text).

\end{document}